\documentclass[fleqn,twoside]{article}
\usepackage{gc}
\usepackage{amsfonts}
\heads{J. Miritzis, P.G.L. Leach and S. Cotsakis}
      {Painlev\'e Integrability of Isotropic Cosmologies}

\begin{document}
\thispagestyle{empty}
\twocolumn[

\Title
{SYMMETRY, SINGULARITIES AND INTEGRABILITY IN COMPLEX \yy
 DYNAMICS IV: \yy PAINLEV\'E INTEGRABILITY OF ISOTROPIC COSMOLOGIES}

\Author
{J. Miritzis\foom 1, P.G.L. Leach$^{2,3}$ and S.
Cotsakis\foom 4} {GEODYSYC, Department of Mathematics, University of the
Aegean, Karlovassi 83 200, Greece}

\Abstract
{We apply the results of singularity analysis to the isotropic cosmological
models in general relativity and string theory with a variety of matter
terms. For some of these models the standard Painlev\'{e} test is sufficient
to demonstrate integrability or nonintegrability in the sense of
Painlev\'{e}.  For others of these models it is necessary to use a less
algorithmic procedure.}


] 
\email 1 {john@env.aegean.gr}
\foox 2 {Permanent address: School of Mathematical and Statistical
        Sciences, University of Natal, Durban, South Africa 4041}
\email 3 {leach@math.aegean.gr}
\email 4 {skot@aegean.gr}

\catcode`\@=11
\renewcommand{\theequation}{\thesection\arabic{equation}}
\@addtoreset{equation}{section}
\catcode`\@=10

\def\dsp{\displaystyle}
\def\d{\mbox{\rm d}}
\def\e{\mbox{\rm e}}


\def\ut#1{{\mathunderaccent\tilde #1}}
\def\cosec{{\rm cosec}}
\def\kd{{\delta _{ij}}}
\def\ke{{\epsilon _{ijk}}}
\def\ha{\mbox{$\frac{1}{2}$}}
\def\tha{\mbox{$\frac{3}{2}$}}
\def\fha{\mbox{$\frac{5}{2}$}}
\def\nha{\mbox{$\frac{n}{2}$}}
\def\oei{\mbox{$\frac{1}{8}$}}
\def\oqr{\mbox{$\frac{1}{4}$}}
\def\tfi{\mbox{$\frac{3}{5}$}}
\def\otw{\mbox{$\frac{1}{12}$}}
\def\oth{\mbox{$\frac{1}{3}$}}
\def\tth{\mbox{$\frac{2}{3}$}}
\def\eth{\mbox{$\frac{8}{3}$}}
\def\osi{\mbox{$\frac{1}{6}$}}
\def\fei{\mbox{$\frac{5}{8}$}}
\def\fra#1#2{\mbox{$\frac{#1}{#2}$}}
\def\f{\frac}
\def\p{\dsp\partial}
\def\arcsinh{{\rm arcsinh}}
\def\arccosh{{\rm arccosh}}
\def\arctanh{{\rm arctanh}}


\def\ga{\alpha}
\def\gb{\beta}
\def\gc{\chi}
\def\gd{\delta}
\def\ge{\epsilon}
\def\gf{\phi}
\def\gg{\gamma}
\def\gi{\iota}
\def\gk{\kappa}
\def\gl{\lambda}
\def\gp{\psi}
\def\gs{\sigma}
\def\gt{\theta}
\def\gu{\upsilon}
\def\gve{\varepsilon}
\def\gw{\omega}
\def\gz{\zeta}
\def\gG{\Gamma}
\def\gD{\Delta}
\def\gF{\Phi}
\def\gL{\Lambda}
\def\gP{\Psi}
\def\gS{\Sigma}
\def\gT{\Theta}
\def\gW{\Omega}

\def\bhw{\hat{\mbox{\boldmath $\omega$}}}
\def\bht{\hat{\mbox{\boldmath $\theta$}}}
\def\bhf{\hat{\mbox{\boldmath $\phi$}}}

\def\bhr{\hat{\mbox{\boldmath $r$}}}
\def\bhL{\hat{\mbox{\boldmath $L$}}}


\def\bfA{{\bf A}}
\def\bfB{{\bf B}}
\def\bfC{{\bf C}}
\def\bfD{{\bf D}}
\def\bfE{{\bf E}}
\def\bfF{{\bf F}}
\def\bfG{{\bf G}}
\def\bfH{{\bf H}}
\def\bfI{{\bf I}}
\def\bfJ{{\bf J}}
\def\bfK{{\bf K}}
\def\bfL{{\bf L}}
\def\bfM{{\bf M}}
\def\bfN{{\bf N}}
\def\bfO{{\bf O}}
\def\bfP{{\bf P}}
\def\bfQ{{\bf Q}}
\def\bfR{{\bf R}}
\def\bfS{{\bf S}}
\def\bfT{{\bf T}}
\def\bfU{{\bf U}}
\def\bfV{{\bf V}}
\def\bfW{{\bf W}}
\def\bfX{{\bf X}}
\def\bfY{{\bf Y}}
\def\bfZ{{\bf Z}}
\def\bfa{{\bf a}}
\def\bfb{{\bf b}}
\def\bfc{{\bf c}}
\def\bfd{{\bf d}}
\def\bfe{{\bf e}}
\def\bff{{\bf f}}
\def\bfg{{\bf g}}
\def\bfh{{\bf h}}
\def\bfi{{\bf i}}
\def\bfj{{\bf j}}
\def\bfk{{\bf k}}
\def\bfl{{\bf l}}
\def\bfm{{\bf m}}
\def\bfn{{\bf n}}
\def\bfo{{\bf o}}
\def\bfp{{\bf p}}
\def\bfq{{\bf q}}
\def\bfr{{\bf r}}
\def\bfs{{\bf s}}
\def\bft{{\bf t}}
\def\bfu{{\bf u}}
\def\bfv{{\bf v}}
\def\bfw{{\bf w}}
\def\bfx{{\bf x}}
\def\bfy{{\bf y}}
\def\bfz{{\bf z}}


\def\p{\dsp \partial}

\def\pb{{\p\over\p b}}
\def\pc{{\p\over\p c}}
\def\pd{{\p\over\p d}}
\def\pe{{\p\over\p e}}
\def\pf{{\p\over\p f}}
\def\pg{{\p\over\p g}}
\def\ph{{\p\over\p h}}
\def\pj{{\p\over\p j}}
\def\pk{{\p\over\p k}}
\def\pl{{\p\over\p l}}
\def\pn{{\p\over\p n}}
\def\po{{\p\over\p o}}
\def\pq{{\p\over\p q}}
\def\pr{{\p\over\p r}}
\def\ps{{\p\over\p s}}
\def\pt{{\p\over\p t}}
\def\pu{{\p\over\p u}}
\def\pv{{\p\over\p v}}
\def\pw{{\p\over\p w}}
\def\px{{\p\over\p x}}
\def\py{{\p\over\p y}}
\def\pz{{\p\over\p z}}

\def\pA{{\p\over\p A}}
\def\pB{{\p\over\p B}}
\def\pC{{\p\over\p C}}
\def\pD{{\p\over\p D}}
\def\pE{{\p\over\p E}}
\def\pF{{\p\over\p F}}
\def\pG{{\p\over\p G}}
\def\pH{{\p\over\p H}}
\def\pI{{\p\over\p I}}
\def\pJ{{\p\over\p J}}
\def\pK{{\p\over\p K}}
\def\pL{{\p\over\p L}}
\def\pM{{\p\over\p M}}
\def\pN{{\p\over\p N}}
\def\pO{{\p\over\p O}}
\def\pQ{{\p\over\p Q}}
\def\pR{{\p\over\p R}}
\def\pS{{\p\over\p S}}
\def\pT{{\p\over\p T}}
\def\pU{{\p\over\p U}}
\def\pV{{\p\over\p V}}
\def\pW{{\p\over\p W}}
\def\pX{{\p\over\p X}}
\def\pY{{\p\over\p Y}}
\def\pZ{{\p\over\p Z}}

\def\da{{\d\over\d a}}
\def\db{{\d\over\d b}}
\def\dc{{\d\over\d c}}
\def\de{{\d\over\d e}}
\def\df{{\d\over\d f}}
\def\dg{{\d\over\d g}}
\def\dh{{\d\over\d h}}
\def\di{{\d\over\d i}}
\def\dj{{\d\over\d j}}
\def\dk{{\d\over\d k}}
\def\dl{{\d\over\d l}}
\def\dn{{\d\over\d n}}
\def\do{{\d\over\d o}}
\def\ddp{{\d\over\d p}}
\def\dq{{\d\over\d q}}
\def\dr{{\d\over\d r}}
\def\ds{{\d\over\d s}}
\def\dt{{\d\over\d t}}
\def\du{{\d\over\d u}}
\def\dv{{\d\over\d v}}
\def\dw{{\d\over\d w}}
\def\dx{{\d\over\d x}}
\def\dy{{\d\over\d y}}
\def\dz{{\d\over\d z}}

\def\dA{{\d\over\d A}}
\def\dB{{\d\over\d B}}
\def\dC{{\d\over\d C}}
\def\dD{{\d\over\d D}}
\def\dE{{\d\over\d E}}
\def\dF{{\d\over\d F}}
\def\dG{{\d\over\d G}}
\def\dH{{\d\over\d H}}
\def\dI{{\d\over\d I}}
\def\dJ{{\d\over\d J}}
\def\dK{{\d\over\d K}}
\def\dL{{\d\over\d L}}
\def\dM{{\d\over\d M}}
\def\dN{{\d\over\d N}}
\def\dO{{\d\over\d O}}
\def\dP{{\d\over\d P}}
\def\dQ{{\d\over\d Q}}
\def\dR{{\d\over\d R}}
\def\dS{{\d\over\d S}}
\def\dT{{\d\over\d T}}
\def\dU{{\d\over\d U}}
\def\dV{{\d\over\d V}}
\def\dW{{\d\over\d W}}
\def\dX{{\d\over\d X}}
\def\dY{{\d\over\d Y}}
\def\dZ{{\d\over\d Z}}


\def\pa#1#2{\f{\dsp \p#1}{\dsp \p#2}}
\def\pp#1#2{\f{\dsp \p^2 #1}{\dsp \p #2^2}}
\def\pam#1#2#3{\f{\dsp \p^2 #1}{\dsp \p #2 \p #3}}

\def\up#1{\p_{#1}}
\def\upr{\p_r}
\def\upt{\p_t}
\def\upu{\p_u}
\def\upv{\p_v}
\def\upw{\p_w}
\def\upx{\p_x}
\def\upy{\p_y}
\def\upz{\p_z}
\def\dm#1#2{\f{\dsp \d #1}{\dsp \d #2}}
\def\dd#1#2{\f{\dsp \d^2 #1}{\dsp \d #2^2}}

\def\dddot#1{\mathinner{\buildrel\vbox{\kern5pt\hbox{...}}\over{#1}}}


\def\be{\begin{equation}}
\def\ee{\end{equation}}
\def\bq{\begin{eqnarray}}
\def\beq{\begin{eqnarray*}}
\def\eeq{\end{eqnarray*}}
\newtheorem{theorem}{Theorem}[section]
\def\ext#1#2{G^{[#1]} #2_{|_{\!\!|_{#2=0}}} = 0}
\def\exta#1{G^{[#1]}}
\def\extb#1#2{G^{[#1]} #2_{|_{\!\!|_{#2=0}}} }
\def\enter{$\longleftarrow\!\!\!^|$}

\def\pain{Painlev\'{e}}
\def\rps{right \pain\ series}
\def\lps{left \pain\ series}
\def\lob{leading order behaviour}


\def\ie{{\it i.e. }}
\def\cf{{\it cf. }}
\def\viz{{\it viz. }}
\def\etal{{\it et al. }}
\def\n{\nonumber}
\def\({\left (}
\def\){\right )}
\def\bi{\begin{itemize}}
\def\ei{\end{itemize}}
\def\z{\eql}
\def\lb{\left[ }
\def\rb{\right] }

\def\pic#1#2#3{\begin{center}\input #1.tex #2
    \end{center}\begin{center}{\it #3}\end{center}}

\def\re#1{(\ref{#1})}
\def\ce{Chazy equation}
\def\gce{generalised Chazy equation}

\section{Introduction}

In the modelling of various phenomena in relativistic cosmology
the final result of the modelling process is a system of ordinary
differential equations, whose solution is by no means obvious. Generally
these systems are of the first order and autonomous, so that, if the system
is two-dimensional, one is assured of the existence of solutions under mild
conditions on the terms in the equations. For systems of dimension three or
more the question of integrability or nonintegrability is extended by the
possibility of chaotic behaviour in the general solution (for a recent
review of the relevant dynamical systems methods in cosmology see
\cite{wa-el}).

One method to determine the integrability of a system is by a performance of
the so-called {\em singularity, or Painlev\'e, analysis} in an effort to
examine whether or not there exists a Laurent expansion of the solution
about a movable pole which contains the number of arbitrary constants
necessary for a general solution. Any other singularities are not permitted
except in the case of branch point singularities which give rise to what is
called the weak Painlev\'e property. A system which is integrable in
the sense of Painlev\'e has its general solution analytic except at
the polelike singularity.

The singularity analysis of ordinary differential equations goes back to the
end of the nineteenth century to the pioneering works of Kowalevskaya
\cite{Kowalevskaya}, Painlev\'e \cite{Painlev1, Painlev2, Painlev3} and
others \cite{others1, others2, others3} and has enjoyed a resurgence of
interest in the last thirty years because of the growing attention paid to
nonlinear equations, both partial and ordinary.  The practical use of the
singularity analysis has been greatly enhanced by the development of the ARS
algorithm \cite{Ablowitz1, Ablowitz2, Ablowitz3}, the simplifications
proposed by Kruskal \cite{Kruskal1, Kruskal2} and the review by Ramani,
Grammaticos and Bountis \cite{Bountis}.  This practical approach  has at
times led to misinterpretations at the hands of practitioners.  The works
of Conte \cite{Conte1, Conte2} have pointed the way to a rigorous
application of the singularity analysis.

In this paper, which is the fourth in a series
\cite{Flessas,II,III} devoted to the investigation of connections
between the three main topics in dynamics namely, symmetry,
singularities and integrability, the topics which superficially
are unrelated in their mathematics and yet are intimately
intertwined, we apply the techniques of singularity analysis to a
number of model systems which have been derived for different
problems in cosmology. The problems which we consider represent
three classes, namely, those of fluid models in general
relativity, scalar field models coupled to fluid cosmologies again
in general relativity and lastly isotropic string cosmologies.
(For more information about the physical characteristics of the
models treated below the reader is urged to consult the cited
papers.) More specifically, we consider the following systems:

1. The one-fluid FRW model in general relativity \cite{Wa} (p. 122):
\begin{eqnarray}
    \dot{H} \z - (1+q)H,\n\\
    \dot{\Omega} \z - 2q (1-\Omega),\qquad
            q = \ha (3\gamma - 2)\Omega\label{1.1}
\end{eqnarray}
the latter of which is, of course, equivalent to the scalar equation
\begin{equation}
    \dot{\Omega} = (2-3\gamma) (1-\Omega)\Omega.\label{1.1a}
\end{equation}
Here, $H$ is the Hubble variable, $\Omega$ the density parameter,
$q$ the deceleration parameter and derivatives are taken with
respect to a dimensionless time parameter $\tau$ defined so that
$dt/d\tau =1/H$, $t$ being the usual time variable, giving the
space sections of constant curvature in these models. As usual, the
parameter $\gamma\in [0,2]$ describes a barotropic fluid.
Although this is the simplest cosmological system, it is included
here for completeness and comparison with more sophisticated ones
introduced below.

2. The two-fluid FRW model in general relativity \cite{Wa} (p. 126):
\begin{eqnarray}
    \dot{\Omega} \z -\ha (b-x)\cos 2\Omega\cos\Omega,\n\\
    \dot{\chi} \z   (1-\chi^2)\sin\Omega,           \label{1.7}
\end{eqnarray}
where the so-called compactified density parameter $\Omega\in
[-\ha\pi,\ha\pi]$ and the transition variable $\chi\in [- 1,1]$ defined in
the above reference are used to describe which fluid is dominant
dynamically. Here  $b>-1 $.

3. The flat one-fluid FRW space-time with a scalar field $\phi $
with an exponential potential in general relativity
\cite{gr-qc/9711068} which reduces to the two-dimensional system
\begin{eqnarray}
    \dot{x} \z - 3x+ 3\(1-\ha\gamma\)x^3-\tha\gamma
                xy^2+\(\tha\)^{1/2}\lambda y^2,\n\\
    \dot{y} \z 3 \(1-\ha\gamma\)x^2y-\tha\gamma y^3
            -\(\tha\)^{3/2}\lambda xy,      \label{1.2}
\end{eqnarray}
where differentiation is again with respect to a dimensionless time
variable. Here $\lambda$ is a positive constant appearing in the scalar
field potential and $\gamma\in (0,2)$, that is, we exclude the ``extreme"
cases of stiff matter and scalar field coming from the fluid contribution.

4. General-relativistic one-fluid FRW models with $n$ scalar
fields $\phi_i$, $i=1,\ldots, n$ with  exponential potentials
\cite{gr-qc/9911075} described in dimensionless variables $(\Omega
,\Phi ,\Psi )$ by the system (this strictly excludes positively
curved models wherein the variables are not defined)
\begin{eqnarray}
\dot{\Psi}_i \z \Psi_i (q- 2) -\sqrt{\tha}k_i\Phi_i^2,
        \cm i =  1,\ldots, n,           \n\\
\dot{\Phi}_i \z \Phi_i\(q+ 1+\sqrt{\tha}k_i\Psi_i\),\qquad
            i =  1,\ldots, n,           \n\\
\dot{\Omega} \z \Omega (2q- 3\gamma + 2),           \label{1.3}
\end{eqnarray}
where
\begin{equation}
    q = \Half (3\gamma - 2)\Omega +
         2\sum_{i = 1}^{n}\Psi_i^2-\sum_{i = 1}^{n}\Phi_i^2.\label{1.4}
\end{equation}
The variables $(\Phi ,\Psi )$ are defined in terms of the
potential and kinetic energies of the scalar fields and $\gamma\in
[0,2]$.

5. The four-dimensional flat string FRW model with negative central charge
deficit described in the compactifying variables $(\xi ,\eta )$ by the
system \cite{gr-qc/9903095}
\begin{eqnarray}
    \dot{\eta} \z \xi^2\(1-\eta^2\),\n\\
    \dot{\xi} \z \(\sqrt{3} +\eta\xi\)\(1-\xi^2\).      \label{1.5}
\end{eqnarray}

6. The ten-dimensional flat string FRW model in the RR sector and
   with a positive cosmological  constant
   described by the system \cite{gr-qc/9910074}
\begin{eqnarray}
    \dot{x} \z (x+\sqrt{3})(1-x^2-y-z) +\ha z(x-\sqrt{3}),\n\\
    \dot{y} \z 2y[(x+\sqrt{3})(1-x^2-y-z) +\ha
                        z(x-\sqrt{3})],\n\\
    \dot{z} \z 2z[(x+\sqrt{3})(1-x^2-y-z)
                +\ha z (x-\sqrt{3})],\nnn \label{1.6}
\end{eqnarray}
for which the dimensionally reduced (four-dimensional) model is
obtained by placing $y =0$ in \re{1.6}.

Another approach to determining the integrability of sets of differential
equations is the use of Noether's theorem for Lagrangian systems and the Lie
theory of extended groups for differential equations in general.  For the
systems which we study here the Lie symmetry approach is not generally
successful because we need to find generalised or nonlocal symmetries to
supplement the point symmetries which are the easiest to obtain.

For the benefit of the reader we describe in \sect 2 the main
points of the Painlev\'{e} process (for a more detailed albeit
elementary introduction we refer the reader to \cite{Tabor}). This
is then used in \sect 3 to study the singularity features of our
cosmological systems. Conclusions are given in \sect 4.

\section{Methodology of Painlev\'{e} analysis}

The singularity analysis which lies at the basis of the Painlev\'e test as
systematised in the ARS algorithm [1--3]
is a specific form of a more general analysis which examines a set of
differential equations for leading order behaviour and next to leading order
behaviour \cite{Feix2, Feix3}.  The latter analysis does not take into
consideration the necessity for a Laurent expansion about a polelike
singularity (or rational branch point) which is required for the
Painlev\'e test.  The essence of the Painlev\'e test (for a set of
ordinary differential equations which is the only type which we consider
here) is that the solution of an $n $-dimensional set of equations
\begin{equation}
        \dot{x}_i = f_i\(t,x\),\label{3.1}
\end{equation}
where the functions $f_i $ are rational in the dependent variables and
algebraic in the independent variable, can be written as either
\begin{equation}
    x_i (\tau) = \sum_{j = 0}^{\infty}a_j\tau^{-p_i+qj} \label{3.2}
\end{equation}
in the case of a Right Painlev\'e Series or
\begin{equation}
    x_i (\tau) = \sum_{j = 0}^{\infty}a_j\tau^{-p_i-qj} \label{3.3}
\end{equation}
in the case of a Left Painlev\'e Series \cite{Lemmer,Feix3}, where
$\tau =t-t_0 $ and $t_0 $ is arbitrary, the exponents $p_i $ are strictly
positive integers (rational numbers in the case of the so-called weak
Painlev\'e test), the parameter $q $ is a (positive) integer (respectively
rational number) and in the coefficients there are $n- 1 $ arbitrary
constants which, together with $t_0 $, give the required number of
arbitrary constants for the general solution of the system \re{3.1}.

The ARS algorithm provides a mechanistic procedure for the determination of
the leading order behaviour and the resonances which are where the arbitrary
constants of integration arise. The first step is to determine the leading
order behaviour by means of the substitution
\begin{equation}
    x_i = \alpha_i\tau^{-p_i}               \label{3.4}
\end{equation}
into the system \re{3.1} and to assemble all possible patterns of values of
the exponents $p_i,i = 1,n $ from the dominant terms of the system \re{3.1}.
For each of these possible patterns the coefficients $\alpha_i $ are
determined and the substitution
\begin{equation}
    x_i = \alpha_i\tau^{-p_i} +\mu_i\tau^{r-p_i}        \label{3.5}
\end{equation}
is made to determine the ``resonances" $r $ at which the arbitrary
constants, $\mu_i $, are introduced.  The resonances and the arbitrary
constants are determined from the leading order behaviour of the system
\re{3.1} by a linearisation process.  The final step for each particular
pattern of leading order behaviour is to substitute a series, truncated at
the highest resonance, to ensure that there is compatability at the
resonances.  If, for a particular pattern of leading order behaviour,
these conditions are satisfied and the series contains $n- 1 $ arbitrary
constants (the $n $th comes from $t_0 $), that particular pattern is said
to pass the Painlev\'e test.  If this is true for all possible patterns of
leading order behaviour, the set of equations is said to possess the
Painlev\'e property and to be integrable.  We note that the concept of
Painlev\'e integrability means the possession of a Laurent series, \ie, the
solution is analytic except at the polelike (branch point in the case
of the weak property) movable singularities.

The algorithm described in the previous paragraph is not always possible or
easy to implement. The first problem arises when not all of the exponents
obtained in the leading order analysis are strictly positive. In the case
that some or all of the exponents are strictly negative one can perform a
homeographic transformation, which preserves the Painlev\'e property, on the
affected variables so that the exponents will now be strictly positive. The
test can then be continued algorithmically. There is no such consolation in
the case of zero exponents in the leading order behaviour.  The algorithm
fails. In this case it is necessary to make a series substitution
commencing at the leading order behaviour to establish whether or not a
sufficient number of arbitrary constants is introduced.  If these constants
enter at early terms in the series, the task is not too difficult with one
of the symbolic manipulation codes.  If this is not the case, the
task can become impossible for a system of moderate complexity.

Another possibility which can occur and create a problem of interpretation
is that the number of arbitrary constants introduced at the resonances is
insufficient to produce the general solution for the system. This
possibility was already observed by Ince, citing from a paper by Chazy
written in 1909, in 1927 \cite{Ince} (p. 355) and treated in some more
detail in \cite{Cotsakis}. Ince described such a solution as a
``singular" solution. This is in accord with the usual usage of the word
in the case of first order equations. However, it is a little unfortunate
that the word ``singular" has to perform two distinct functions in the
description of the integrability of one system.  In an attempt to separate
the two uses of the word, Cotsakis and Leach \cite{Cotsakis} described this
type of solution as a ``partial" solution confined to a submanifold of the
space of initial conditions, but this has not met universal acceptance.
One is reluctant to use ``particular" solution because of the commonness of
the term in the solution of nonhomogeneous linear equations. In Greek there
are three words, "$\alpha\nu\acute{\omega}\mu\alpha\lambda\eta$",
"$\epsilon\iota\delta\iota\kappa\acute{\eta}$"  and
"$\iota\delta\iota\acute{\alpha}\zeta o\upsilon\sigma\alpha$"
meaning ``singular", ``particular" and ``peculiar" respectively, which in
Greek are used to describe these different types of solution.

In this paper we propose to use the adjective ``peculiar" to describe a
solution of the type given by Ince. As far as we are concerned here, the
question is whether or not the existence of such solutions violates the
requirement that all possible solutions pass the Painlev\'e test, which, one
must recall, requires the existence of the correct number of arbitrary
constants in the Laurent expansion for each pattern of leading order
behaviour. According to Tabor \cite{Tabor} (p. 330), this is necessary for
the possession of the Painlev\'e property. However, a recent paper
\cite{Flessas} has demonstrated integrability in the case that of the two
possible patterns of leading order behaviour one satisfied the Painlev\'e
test and the other was lacking one arbitrary constant. This demonstration
was for a single example and one would want a sounder basis for making a
definite claim about integrability in cases for which the requirements of
the Painlev\'e property were only partially satisfied for some patterns of
leading order behaviour.

Finally we recall that the possession of the Painlev\'e Property is
representation dependent and is definitely preserved only under a
homeographic transformation.  However, there are times when we can introduce
a transformation with profit such as in the system described by \re{1.7}.
The nature of the integrability of the original system in comparison with
that of the transformed system will be tempered by the nature of the
particular transformation.

\section{Applications of Painlev\'e analysis to models}

\subsection{One-fluid FRW model, \re{1.1}}

In the case of \re{1.1} it is useful to introduce the rescaling transformation
\begin{equation}
H \longrightarrow x,\quad \Omega \longrightarrow y,\quad t \longrightarrow
\ha (3\gamma - 2)t\label{2.1.1}
\end{equation}
to give the system
\begin{eqnarray}
    \dot{x} \z - (\sigma +y)x,  \n\\
    \dot{y} \z - 2y (1-y),      \label{2.1.2}
\end{eqnarray}
where $\sigma =\ha (3\gamma - 2) $.
When we make the usual substitution for the leading order behaviour, we obtain
\begin{eqnarray}
    \alpha p\tau^{p-1} \z -\alpha\beta\tau^{p+q}\n\\
    \beta q\tau^{q-1} \z  2\beta^2\tau^{2q}.    \label{2.1.3}
\end{eqnarray}
From the second \eq \re{2.1.3} it is evident that $q = - 1 $ and $\beta =
-\ha $. From the exponents of the first \eq \re{2.1.3} the value of $p $ is
arbitrary. However, equality of the coefficients of the leading order
powers imposes the requirement that $p =\ha $.  Although the value of $p$
is not strictly negative, we can proceed with the Painlev\'e test without a
further transformation since it is a nonintegral rational number.
We note that the value of $\alpha $ is unspecified.

To obtain the resonances we make the substitution
\begin{eqnarray}
    x \z \alpha\tau^{1/2} +\mu\tau^{r+1/2}, \n\\
    y \z \beta\tau^{- 1} +\nu\tau^{r- 1}    \label{2.1.4}
\end{eqnarray}
to obtain the condition
\begin{equation}
    \left|\begin{array}{ccc}
    r+\ha +\beta &\quad &\alpha\\
    0 & &r- 1-4\beta
            \end{array}\right| = 0\label{2.1.5}
\end{equation}
that there be a nontrivial solution.  The condition \re{2.1.5}
gives the resonances $r = - 1,0 $ and so the system \re{2.1.2}
passes the Painlev\'e test.  Hence the system \re{1.1} possesses
the Painlev\'e property.  The first few terms of the Laurent expansion are
\begin{eqnarray}
    x (\tau) \z a_0 \Bigl\{\tau^{1/2} - (\sigma +\ha)\tau^{3/2}     \n\\
    && \cm + \ha[(\sigma +\ha)^2+\osi]\tau^{5/2} +\ldots \Bigr\},   \n\\
y (\tau) \z -\ha\tau^{- 1} +\ha -\osi\tau^2+\ldots.     \label{2.1.6}
\end{eqnarray}

\subsection{Two-fluid model, \re{1.7}}

The system \re{1.7} is not in a suitable form for applying the Painlev\'e
analysis.  We have two choices for the introduction of new variables.
Interestingly both changes of variables produce essentially the same
results. In the first instance we introduce the new variables
\begin{equation}
    x = \chi\qquad\mbox{\rm and }\qquad y = \cosec\Omega,\label{2.5.1}
\end{equation}
so that the system \re{1.7} becomes
\begin{eqnarray}
    \dot{x}y \z   \(1-x^2\)\n\\
    \dot{y}y^2 \z \ha (b-x)\(y^2-1\)\(y^2-2\).\label{2.5.2}
\end{eqnarray}
The usual leading term analysis gives the exponents $p = - 1 $ and
$q = 0 $, so that the ARS logarithm is not applicable.

We do not continue with the analysis of the system in \re{2.5.2} since,
as we noted above, the second change of variables produces essentially the
same results.

The second transformation, which we consider in detail here, has
the advantage of being one-to-one and continuous over the defined intervals of
the original variables.  We set
\begin{equation}
    x = \chi\qquad \mbox{\rm and} \qquad y = \sin\Omega\label{2.5.50}
\end{equation}
to obtain the system
\begin{eqnarray}
    \dot{x} \z (1- x^2)y\n\\
    \dot{y} \z -\ha (b-x) (1-2y^2) (1-y^2).\label{2.5.51}
\end{eqnarray}
The same exponents are obtained as for \re{2.5.2}. We make the {\it Ansatz}
\begin{equation}
   x = \sum_{i=0}a_i\tau^{i- 1},\cm y = \sum_{i=0} b_i\tau^i\label{2.5.3}
\end{equation}
and substitute this into the system \re{2.5.51} to obtain the pair of
relations
\bearr
   (i- 1)a_i\tau^{i- 2}= -a_ia_jb_k\tau^{i+j+k- 2} +b_i\tau^i,
                        \label{2.5.4}
\ear
\bear
 ib_i\tau^{i- 1} \z -\ha b(1-3b_0ib_j\tau^{i+j} +
            2b_i b_j b_k b_l \tau^{i+j+k+l}) \n\\ \lal\
    +a_i( \tau^{i- 1} - 3b_jb_k\tau^{i+j+k- 1} \n\\ \lal\
        + 2b_jb_kb_lb_m\tau^{i+j+k+l+m - 1}),           \label{2.5.5}
\ear
from which we are able to deduce the first few terms of the expansions for
$x $ and $y $.  From the first two terms of \re{2.5.4} we obtain
\begin{equation}
    a_0b_0 = 1\qquad\mbox{\rm and}\qquad a_1 = - \ha a_0b_1\label{2.5.52}
\end{equation}
and from the first term of \re{2.5.5}
\begin{equation}
    a_0\(1-3b_0^2+ 2b_0^4\) = 0.\label{2.5.53}
\end{equation}
Since $a_0 \neq 0 $, it follows from \re{2.5.53} that $b_0^2 = 1,\ha $.  In
either case the second term of \re{2.5.5} gives $b_1 = 0 $ and consequently
$a_1 = 0 $. The third term of \re{2.5.5} reduces to $b_2 =b_0^2b_2 $, so
that, for the first possibility for the value of $b_0 $, $b_2 $ is arbitrary
and, for the second possibility, zero.  In fact for the second possibility
all subsequent coefficients $b_i $ are zero and we obtain the solution
\begin{eqnarray}
    x (\tau) \z \frac{1}{b_0\tau}
         +\frac{b_0\tau}{3} -\frac{\(b_0\tau\)^3}{45} + \ldots,\n\\
    y (\tau) \z b_0,            \label{2.5.54}
\end{eqnarray}
where $b_0^2 =1/2 $, which is certainly a peculiar solution.

In the case that $b_0^2 = 1 $ we obtain a more standard solution.
We have that $b_2 $ is arbitrary and this provides us with the
second arbitrary constant required for a general solution of the
original system.  We have
\begin{eqnarray}
    x \z a_0\tau^{- 1} +\oth\(b_0-2b_2\)\tau +\oqr bb_0b_2\tau^2+\ldots,\n\\
    y \z b_0+b_2\tau^2 +\ldots,     \label{2.5.6}
\end{eqnarray}
where the coefficients $a_0 $ and $b_0 $ have been given above.

In the expansions we have obtained, only the second one has the required
number of arbitrary constants, and we cannot conclude that the system
\re{1.7} is integrable in the sense of Painlev\'e.  However, we do note that
it is possible to obtain a first integral of the original system \re{1.7}
and reduce the solution to a rather complicated quadrature. In fact the
system can be written in terms of a Lagrangian and is Hamiltonian, so that
the existence of the first integral immediately guarantees integrability in
the sense of Liouville.

\subsection{Flat FRW with one fluid and an exponential potential, \re{1.2}}

Before we begin the singularity analysis of \re{1.2}, it is appropriate to
simplify the system by the rescaling
\begin{eqnarray}
       t &\longrightarrow &\frac{2\(2-\gamma\)}{3\lambda^2}t, \n\\
       x &\longrightarrow &\frac{\lambda\sqrt{3}}{\sqrt{2}\(2-\gamma\)}x,\n\\
       y &\longrightarrow &\beta y\quad\mbox{\rm with}\quad \beta^2
           =\frac{3\lambda^2}{2\gamma \(2 -\gamma\)},\label{2.2.1}
\end{eqnarray}
in which the sign of $\beta $ may be taken as positive without loss of
generality.  The system \re{1.2} now has the simpler appearance
\begin{eqnarray}
    \dot{x} \z - Ax + By^2+ x^3-xy^2\n\\
    \dot{y} \z -xy +x^2y-y^3,\label{2.2.2}
\end{eqnarray}
where $A = 8\(1-\ha\gamma\)/\(3\lambda^2\) $ and $B =\lambda^2/\(\gamma
(2-\gamma)\) $.

We determine the leading order behaviour of the system \re{2.2.2}
to be $x (\tau ) =\alpha\tau^{-1/2} $ and $y(\tau)=\beta\tau^{-1/2} $ with
the constraint $\alpha^2-\beta^2 =-1/2$.  We find that the resonances are at
$r = - 1,0 $, where the first resonance is generic and the second indicates
that one of the coefficients of the leading order behaviour is arbitrary,
which is in accordance with the above constraint.  If we make the
substitutions
\begin{eqnarray}
    x (\tau) \z \sum_{i = 0} a_i\tau^{(i- 1)/2},\n\\
    y (\tau) \z \sum_{i = 0} b_i\tau^{(i- 1)/2},\label{2.2.3}
\end{eqnarray}
we find that the first few terms of the expansion are given by
\begin{eqnarray}
 a_0 \z \mbox{\rm arbitrary}\n\\
 a_1 \z B+\tth(2+5B)a_0^2+\eth(1+B)a_0^4\n\\
 a_2 \z \osi a_0(3B(1 + 5B) + 8 (1+ 2B (4+ 5B))a_0^2\n\\
 && + 28 (1+B) (2+5B)a_0^4\n\\
 &&+ 80 (1+B)^2a_0^6-6A  (1+a_0^2)\n\\
b_0^2 \z a_0^2 +\ha\n\\ b_1 \z \tth a_0 (- 1+ 2B+ 4(1+B)a_0^2)\n\\
        b_2 \z \osi b_0 (3 (- 1+B)B\n\\
 &&+ 2 (- 2-3A + 2B+ 16B^2 )a_0^2\n\\
 && + 4 (1+B)(2+ 23B)a_0^4\n\\&&+ 80 (1+B)^2 a_0^2 ).         \label{2.2.4}
\end{eqnarray}
We conclude that the system \re{1.2} is integrable in the sense of Painlev\'e.

\subsection{Exponential potential with one fluid \re{1.3}}

We consider the case in which $n = 1 $.  When we make the
substitution for $q $ in the system \re{1.3}, we obtain
\begin{eqnarray}
    \dot{\Psi} \z \ha (3\gamma - 2)\Psi\Omega + 2\Psi^3-\Phi^2\Psi -
                    2\Psi-\sqrt{\tha} K\Phi^2,\n\\
    \dot{\Phi} \z \ha (3\gamma - 2)\Phi\Omega + 2\Psi^2\Phi -\Phi^3
                    +\Phi +\sqrt{\tha} K\Psi\Phi,\n\\
    \dot{\Omega} \z (3\gamma - 2)\Omega^2+ 4\Psi^2\Omega
        - 2\Phi^2\Omega - ( 3\gamma - 2)\Omega. \label{2.3.1} \nnn
\end{eqnarray}
We find that the leading order behaviour is given by
\bearr
    \Psi (\tau) = a\tau^{- 1/2}, \quad
    \Phi (\tau) = b\tau^{- 1/2}, \quad
    \Omega(\tau) = c\tau^{- 1}, \label{2.3.2}
\ear
subject to the constraint that
\begin{equation}
(3\gamma - 2)c+ 4a^2-2b^2 = - 1.\label{2.3.3}
\end{equation}
The analysis of the dominant terms for the
resonances is facilitated by the substitutions
\begin{equation}
x = 4\Psi^2,\quad y = - 2\Phi^2,\quad z = (3\gamma - 2)\Omega.\label{2.3.4}
\end{equation}
(Note that this transformation does not preserve the Painlev\'e property.
However, it is satisfactory for the purposes of this immediate analysis.)

The dominant terms in the system \re{2.3.1} are now
\begin{eqnarray}
    \dot{x} \z x (x+y+z)\n\\
    \dot{y} \z y (x+y+z)\n\\
    \dot{z} \z z (x+y+z).\label{2.3.5}
\end{eqnarray}
The leading order behaviour of \re{2.3.5} is given by
\begin{eqnarray}
    x \z \alpha\tau^{- 1},\n\\ y \z \beta\tau^{- 1},\n\\ z \z
                    \gamma\tau^{- 1},    \label{2.3.6}
\end{eqnarray}
subject to the constraint $\alpha +\beta +\gamma = - 1 $.
(The constant $\gamma $ in \re{2.3.6} is not to be confused with the physical
 constant in the original system.)
 We determine the resonances by substituting into \re{2.3.5}
\begin{eqnarray}
    x \z \alpha\tau^{- 1} +\mu\tau^{s- 1},\n\\ y \z \beta\tau^{- 1}
    +\nu\tau^{s- 1},\n\\ z \z \gamma\tau^{- 1} +\rho\tau^{s- 1}
        \label{2.3.7}
\end{eqnarray}
to obtain the linearised system
\begin{equation}
\lb\begin{array}{ccccc}
    s-\alpha &\quad & -\alpha & & -\alpha\\
    -\beta& &s-\beta &\quad & -\beta\\
    -\gamma & & -\gamma & &s-\gamma
                \end{array}\rb \lb\begin{array}{c}
    \mu\\ \nu\\  \rho
                \end{array}\rb = 0 \label{2.3.8}
\end{equation}
which has a nontrivial solution if $s = - 1,0 (2) $.  Thus we see
that there is a double zero resonance which is consistent with the
constraint.  These results pass over to the original system and,
since two arbitrary constants enter at the leading order terms,
the system \re{2.3.1} passes the Painlev\'e test for this pattern
of leading order behaviour.

We present the first few terms of the Laurent expansion of the original
system \re{2.3.5}.
\begin{eqnarray}
    \Psi \z a_0\tau^{-\ha} +a_1+a_2\tau^{\ha} +\ldots,\n\\
    \Phi \z b_0\tau^{-\tau} +b_1+b_2\tau^{\ha} +\ldots,\n\\
    \Omega \z c_0\tau^{- 1} +c_1\tau^{-\tau} +c_2+\ldots,\label{2.3.9}
\end{eqnarray}
where
\begin{eqnarray}
a_{1} \z -\sqrt{6}k\left( 1+4a_{0}^{2}\right) b_{0}^{2}\n\\ a_{2}
\z \oqr a_{0}[ -8+A-2\left( 2+A+12k^{2}\right)
b_{0}^{2}\n\\&&+180k^{2}b_{0}^{4}]\n\\ && +a_{0}^{3}\left[ \left(
-4+A+18k^{2}b_{0}^{2}\right) \left(
 -1+10b_{0}^{2}\right) \right] \n\\
b_{1} \z \sqrt{6}ka_{0}b_{0}\left( 1-4b_{0}^{2}\right) \n\\ b_{2}
\z \oqr b_{0}[
4+A-16a_{0}^{2}+4Aa_{0}^{2}+12k^{2}a_{0}^{2}\n\\&&-2 ( 2+A+6k^{2}(
1+18a_{0}^{2}) ) b_{0}^{2}\n\\ &&+36k^{2}\left(
1+20a_{0}^{2}\right) b_{0}^{4}] \n\\ c_{0} \z \frac{1}{A}\left(
-1-4a_{0}^{2}+2b_{0}^{2}\right) \n\\ c_{1} \z \frac{1}{A}[
8\sqrt{6}ka_{0}b_{0}^{2}\left( 1+4a_{0}^{2}-2b_{0}^{2}\right) ]
\n\\ c_{2} \z -\frac{1}{2A}[ (1+4a_{0}^{2}-2b_{0}^{2})(
-A\n\\&&-16a_{0}^{2}+4Aa_{0}^{2}-2( 2+A+36k^{2}a_{0}^{2})
b_{0}^{2}\n\\ &&+12k^{2}\left( 3+76a_{0}^{2}\right) b_{0}^{4}) ]
,\label{2.3.20}
\end{eqnarray}
in which we have taken $a_0 $ and $b_0 $ as two arbitrary constants and
expressed the third, $c_0 $, in terms of them using the constraint \re{2.3.3}.

We note that the system \re{2.3.5} is easily integrated to give
the closed-form solution
\begin{eqnarray}
    x\left( t\right) \z -\frac{1}{\left( 1+c_{1}+c_{2}\right) \left(
            t-c_{3}\right) } \n\\
    y\left( t\right) \z -\frac{c_{2}}{\left( 1+c_{1}+c_{2}\right) \left(
            t-c_{3}\right) } \n\\
    z\left( t\right) \z -\frac{c_{1}}{\left( 1+c_{1}+c_{2}\right) \left(
                t-c_{3}\right) }.\label{2.3.21}
\end{eqnarray}
Hence the leading order analysis yields the exact solution of the system
consisting of the dominant terms only.

Apart from the leading order behaviour given in \re{2.3.6}, there are other
possibilities.  Denoting the leading order powers of $x $, $y $ and $z $ by
$p $, $q $ and $r $, we have the following possibilities apart from that in
\re{2.3.6} listed in the table:
$$
    \begin{array}{|r|r|r|}
    \hline p &q &r \\ \hline - 1 & - 1 & \geq 0\\ \geq 0 & - 1 & - 1\\
    - 1 & \geq 0 & - 1\\ - 1 & \geq 0 & \geq 0\\ \geq 0 & - 1 & \geq
        0\\ \geq 0 & \geq 0 & - 1\\ \hline
\end{array}.
$$
In each of these cases it is necessary to make a series substitution since
the ARS algorithm for the application of the Painlev\'e test is no longer
appropriate. We summarise the results:

1. We substitute the series
\begin{eqnarray}
    \Psi \z \sum_{i=0}a_i\tau^{(i-1)/2},\quad \Phi =
                \sum_{i=0}b_i\tau^{(i-1)/2},\n\\
    \Omega \z \sum_{i=0}c_i\tau^{i/2}\label{20.1}
\end{eqnarray}
to find that $a_0$ is arbitrary and $c_0=c_1=c_2=\ldots=0$, so that we have
a peculiar solution with two arbitrary constants ($a_0$ and $t_0$) and one
of the functions, $c(\tau)$,  having only the trivial solution.

2. The series substituted are now
\begin{eqnarray}
 \Psi \z \sum_{i=0}a_i\tau^{i/2},\quad \Phi =
 \sum_{i=0}b_i\tau^{(i-1)/2},\n\\
\Omega \z \sum_{i=0}c_i\tau^{(i-2)/2}.\label{20.2}
\end{eqnarray}
We find that $a_0$ and $c_0$ are arbitrary and that $b_0 =
\sqrt{(1+Ac_0)/2}$, so that the series solutions do contain three arbitrary
constants when $t_0$ is included.

3. In this case we put
\begin{eqnarray}
 \Psi \z \sum_{i=0}a_i\tau^{(i-1)/2},\quad \Phi =
 \sum_{i=0}b_i\tau^{i/2},\n\\
\Omega \z \sum_{i=0}c_i\tau^{(i-2)/2}.\label{20.3}
\end{eqnarray}
We distinguish three subcases.

Subcase (a):  the coefficients $a_0=a_1=a_2=\ldots=0$ and
$b_0=b_1=b_2=\ldots=0$ and $c_0=-1/A$ indicate that we have trivial
solutions for $\Psi$ and $\Phi$ and a nontrivial series with arbitrary
constant $t_0$ for $\Omega$.

Subcase (b): we find that $a_0$ is arbitrary and $b_0=b_1=b_2=\ldots=0$, so
that we have the trivial solution for $\Phi$ and a two parameter solution
($a_0$ and $t_0$) for $\Psi$ and $\Omega$.

Subcase (c): In this subcase $a_0=\ha i$, $b_0=b_1=b_2=\ldots=0$
and $c_0=c_1=c_2=\ldots=0$, so that we have trivial solutions
for $\Phi$ and $\Omega$ and a one-parameter solution for $\Psi$.

All three subcases present us with peculiar solutions in the form of series.

4. With the substitution
\begin{eqnarray}
 \Psi \z \sum_{i=0}a_i\tau^{(i-1)/2},\quad \Phi =
 \sum_{i=0}b_i\tau^{i/2},\n\\
\Omega \z \sum_{i=0}c_i\tau^{i/2}\label{20.4}
\end{eqnarray}
we find that $a_0=0$, so that there is no singular behaviour in any of the
series representations of the three functions.  However, the constants
$a_1$, $b_0$ and $c_0$, the leading terms of each of the series, are
arbitrary, and so we obtain a three-parameter representation of the solution.

5. When we make the substitution
\begin{eqnarray}
 \Psi \z \sum_{i=0}a_i\tau^{(i-1)/2},\quad \Phi =
 \sum_{i=0}b_i\tau^{(i-1)/2},\n\\
\Omega \z \sum_{i=0}c_i\tau^{i/2},\label{20.1}
\end{eqnarray}
we find two subcases.

Subcase (a): the coefficient $b_0=0$, which removes any possible singular
behaviour, and the coefficients $a_0$, $b_1$ and $c_0$ are arbitrary,
thereby providing a three-parameter series representation for $\Psi$, $\Phi$
and $\Omega$.

 Subcase (b): the coefficient $b_0 = 2^{-1/2}$.  All coefficients $c_i$ are
 zero and all coefficients $a_i$ and $b_i,\ i > 0$ are expressed in terms of
the parameter $A$ of the system.  Thus we have a trivial solution for
$\Omega$ and a one-parameter solution (the location of the singularity
$t_0$ of $\Phi$) for the two functions $\Psi$ and $\Phi$.

The first subcase presents a solution without a singularity and the second
one a peculiar solution with singularity.

6. For the last case we substitute
\begin{eqnarray}
 \Psi \z \sum_{i=0}a_i\tau^{i/2},\quad \Phi =
 \sum_{i=0}b_i\tau^{i/2},\n\\
\Omega \z \sum_{i=0}c_i\tau^{(i-2)/2}.\label{20.6}
\end{eqnarray}
We find that $c_0=c_1=0$, thereby removing any possible singular behaviour,
and that $a_0$, $b_0$ and $c_2$ are arbitrary.  Thus we have a
three-parameter series representation of the solutions for $\Psi$, $\Phi$
and $\Omega$.

Altogether the results support the proposition that the system
\re{1.3} is integrable.

\subsection{Four-dimensional flat string FRW, \re{1.5}}

In the analysis of the leading order behaviour of the system
\re{1.5} we find the leading order behaviour
\begin{equation}
    \eta = \alpha\tau^{- 1/3},\quad\ \xi = \beta\tau^{- 1/3},\quad\
        3\alpha\beta^2 = 1.\label{2.4.1}
\end{equation}
In accordance with the constraint in \re{2.4.1} we find that the resonances
are at $r = - 1,0 $.  Since the second arbitrary constant enters at the
leading order terms, the full system, \re{1.5}, passes the ``weak"
Painlev\'e test and so is integrable in the sense of Painlev\'e.  The first
few terms of the Laurent expansion are
\begin{eqnarray}
    \eta \z \frac{1}{3b_{0}^{2}\tau ^{1/3}}+\frac{-1+3\sqrt{3}
          b_{0}^{3}+18b_{0}^{6}}{15b_{0}^{4}}\tau ^{1/3}\n\\&&+\ldots \\
    \xi \z \frac{b_{0}}{\tau ^{1/3}}-\frac{3\left( -1+3\sqrt{3}
        b_{0}^{3}+3b_{0}^{6}\right) }{15b_{0}^{4}}\tau ^{1/3}
                \n\\&&
                +\ldots.\label{2.4.2}
\end{eqnarray}

\subsection{10-dimensional flat string FRW, \re{1.6}}

For the system \re{1.6}
\begin{eqnarray}
\dot{x} \z ( x+\sqrt{3}) ( 1-x^{2}-y-z) +\half z ( x-\sqrt{3}),  \\
\dot{y} \z 2y\left[ ( 1-x^{2}-y-z) +\half z\right],  \\
\dot{z} \z 2z \left[( 1-x^{2}-y-z) -\half(1-z-\sqrt{3}x)\right]\label{2.8.1}
\end{eqnarray}
we determine the leading order behaviour of this system to be $x\left( \tau
\right) =\alpha \tau ^{-1/2},$ $y\left( \tau \right) =\beta \tau ^{-1}$ and
$z\left( \tau \right) =\gamma \tau ^{-1}$ with the constraint $2\alpha
^{2}+2\beta +\gamma =1.$ We determine the resonances by the substitution of
\begin{eqnarray}
    x( \tau ) \z \alpha \tau ^{-1/2}+\mu \tau ^{s-1/2},\n\\
    y( \tau ) \z \beta \tau ^{-1}+\nu \tau  ^{s-1},\n\\
    z( \tau ) \z \gamma \tau ^{-1}+\rho \tau^{s-1}      \label{2.8.2}
\end{eqnarray}
into \re{1.6} to obtain the linearised system
\begin{equation}
    \left(
    \begin{array}{ccc}
    s+2\alpha ^{2} & \alpha  & \alpha /2 \\
    4\alpha \beta  & s+2\beta  & \beta  \\
    4\alpha \gamma  & 2\gamma  & s+\gamma
    \end{array}
    \right) \left(
    \begin{array}{c}
    \mu  \\
    \nu  \\
    \rho
    \end{array}
    \right) =0\label{2.8.3}
\end{equation}
which has a nontrivial solution if $s=-1,0\left( 2\right) .$ There is a double
zero resonance which is consistent with the constraint and the system passes
the Painlev\'e test for this pattern of leading order behaviour. The first few
terms of the Laurent expansion
\begin{eqnarray}
x\left( \tau \right)  \eql a_{0}\tau^{-1/2}+a_{1}+a_{2}\tau ^{1/2}+\ldots \n\\
y\left( \tau \right)  \eql b_{0}\tau^{-1}+b_{1}\tau ^{-1/2}+b_{2}+b_{3}\tau
        ^{1/2}\ldots \n\\
z\left( \tau \right)  \eql c_{0}\tau^{-1}+c_{1}\tau ^{-1/2}+c_{2}+c_{3}\tau
        ^{1/2}\ldots\label{2.8.4}
\end{eqnarray}
are given by
\begin{eqnarray}
a_{1} \eql \left( 1/\sqrt{3}\right) [ 3\left( -3+4b_{0}\right)
\n\\&&-2a_{0}^{2}\left( -11+6a_{0}^{2}+6b_{0}\right) ]  \n\\ a_{2}
\eql \left( a_{0}/2\right) [-93+180a_{0}^{6}+4b_{0}\left(
68-45b_{0}\right) \n\\&&+6a_{0}^{4}\left( -77+60b_{0}\right) \n\\
&&+2a_{0}^{2}\left( 188+3b_{0}\left( -107+30b_{0}\right) \right) ]
\n\\ b_{1} \eql -\left( 4/\sqrt{3}\right) a_{0}b_{0}\left(
-5+6a_{0}^{2}+6b_{0}\right) \n\\ b_{2} \eql \left( b_{0}/3\right) [
-63+168b_{0}+2( 342a_{0}^{6}-54b_{0}^{2}\n\\&&+3a_{0}^{4}\left(
-199+228b_{0}\right)\n\\&& +a_{0}^{2}\left( 290+3b_{0}\left(
-217+114b_{0}\right) \right) ) ]  \n\\ c_{0}
\eql 1-2a_{0}^{2}-2b_{0} \n\\ c_{1} \eql \left( 2/\sqrt{3}\right)
a_{0}\left( -1+2a_{0}^{2}+2b_{0}\right) (
-13\n\\&&+12a_{0}^{2}+12b_{0}) \n\\ c_{2} \eql \left( -1/3\right)
\left( -1+2a_{0}^{2}+2b_{0}\right) [
684a_{0}^{6}\n\\&&+6a_{0}^{4}\left( -229+228b_{0}\right)\n\\ &&
+2a_{0}^{2}\left( 392+57b_{0}\left( -13+6b_{0}\right) \right)
\n\\&&-3\left( 31+4b_{0}\left( -17+9b_{0}\right) \right) ]
.\label{2.8.5}
\end{eqnarray}

Note that the constraint is apparent in the first of the relations for the
coefficients $c_i$.

\section{Discussion}

The examples discussed in this paper may prove useful when one is interested
in having a theory to  decide the question of what is the general
singularity pattern of isotropic cosmologies in different gravity theories.
This question, apart from its purely mathematical interest, is believed to
be related to recent observations of the possible oscillatory nature of
the universe, exemplified in an oscillatory behaviour in the Hubble
parameter \cite{br-etal}.

 Among the models discussed in this paper, the two-fluid FRW model in
 general relativity presents the most interesting dynamical behaviour. All
other models are either strictly integrable in the sense of possessing the
strong Painlev\'e property,  or they have branch-point singularities
indicating weak integrability in the sense of Painlev\'e.

The two-fluid model possesses the interesting property of what can be called
{\em singular envelopes}, first discussed by Ince and rediscovered in a
different context in \cite{Cotsakis}. That is, although the general solution
is unknown, any possible nonintegrable or even chaotic behaviour may be
confined to a region of phase space enveloped by the peculiar solutions
in the sense given at the end of the previous section. It is interesting to
ask whether this property is rigid, that is, is it maintained when two-fluid
models are considered either in other gravity theories, or in more
general Bianchi models in general relativity. The former question is
currently under investigation and can be reformulated as a two-fluid plus a
scalar field model in the Einstein frame.

Another question is whether integrability in the case of scalar field
models is maintained when one considers more general potentials. This  is
known not to be the case even for a general FRW model with a scalar field
that has a simple quadratic potential \cite{co-ho}. One would like to be
able to relate the integrability properties of different cosmological
spacetimes in the context of different gravity theories and matter fields in
an effort to understand the significance of exceptional non-integrable cases
as opposed to generic, integrable ones in the simple frame of isotropic
models before one moves on to more difficult homogeneous but anisotropic
case. Problems in this direction are currently under investigation
\cite{co-le-mi}.

\Acknow
{PGLL thanks the Director of GEODYSYC, Dr S Cotsakis, and the Department of
Mathematics, University of the Aegean, for the use of their facilities
during the time when this work was performed and acknowledges
the continuing support of the National Research Foundation of
South Africa and the University of Natal.}

\end{document}